\documentstyle[prb,aps,twocolumn,graphicx]{revtex}
\draft
\begin{document}
\newcommand{\mgb}{MgB$_2$}
\draft
\title{Theoretical de Haas-van Alphen Data and Plasma Frequencies 
of MgB$_2$ and TaB$_2$}
\author{S.\ Elgazzar$^1$, P.M.\ Oppeneer, S.-L.\ Drechsler, R.\ Hayn, 
H.\ Rosner$^\dag$}
\address{Institut f\"{u}r Festk\"{o}rper- und Werkstofforschung,
P.O.\ Box 270116, D-01171 Dresden, Germany}
\address{$^\dag$Dept.\ of Physics, University of California, Davis CA 95616, USA}
\maketitle
\indent
\begin{minipage}[]{16.2cm}
{\small
The de Haas-van Alphen-frequencies as well as the effective masses for 
a magnetic field parallel to the crystallographic $c$-axis are
calculated within the local spin density approximation (LSDA) for MgB$_2$ and 
TaB$_2$. In addition, we analyze the plasma
frequencies computed for each Fermi surface sheet. We find a large anisotropy 
of Fermi velocities in MgB$_2$ in 
difference to the nearly isotropic behavior in TaB$_2$. 
We compare calculations performed within the relativistic non-full
potential augmented-spherical-wave (ASW) scheme and the
scalar-relativistic 
full potential local orbital (FPLO) scheme.
A
significant dependence for small cross sections on the bandstructure
method is found. The comparison 
with the first available experimental de Haas-van Alphen-data by 
Yelland {\it et al.} 
(Ref.\ 19) shows deviations from the electronic structure 
calculated 
within both  
L(S)DA approaches although the cross section predicted by FPLO are 
closer to the experimental data. The elucidation of the relevant 
many-body effects beyond the
standard LDA is 
considered as a possible 
key problem to understand the superconductivity in MgB$_2$.}
\end{minipage}\hfill 
\indent
\narrowtext
\section{Introduction}
The unexpected discovery of superconductivity in
MgB$_2$ \cite{Nagamatsu}, with a surprisingly high transition
temperature $T_{c}\sim$ 40 K, has raised considerable interest in the
clarification
 of its electronic structure and its relationship to the mechanism of 
superconductivity. The present theoretical approaches to this challenging
problem can be divided roughly into three scenarios emphasizing: (i) a 
conventional electronic structure and a more or less standard
electron-phonon interaction within a (multiple) wide-band model \cite{An}$^-$\cite{Rosnern}, 
(ii) the presence of strong electron-electron \cite{hirsch} 
or magnetic interactions \cite{imada}, and (iii)
strong nonadiabatic \cite{pietronero} or polaronic \cite{alexandrov} effects.
In the present paper we shall mainly consider the first approach which 
is based on calculations of the  electronic structure within the 
local (spin) density approximation (L(S)DA) to the density-functional
theory.
The L(S)DA 
 band structure calculations  \cite{An}$^-$\cite{Rosnern} revealed
four 
sheets of the Fermi surface of  MgB$_2$ with a remarkable high
anisotropy of Fermi velocities for $\sigma$-holes. That anisotropy is
important to 
explain the significant anisotropy of the upper critical field $H_{c2}(T)$ in
terms 
of a two-band Eliashberg model \cite{Fuchs}$^-$\cite{Rosnern}. 
The relatively high value 
of $H_{c2}(0)$ 
is related to the strong electron-phonon interaction  of the
$\sigma$-holes on the tube-like Fermi surface sheets \cite{Rosnern}. 
Details of the  
electronic structure are important for a precise understanding of
the pairing mechanism, like, e.g., by comparing the calculated ``undressed''
electronic effective  
mass with its actual dressed value due to the electron-phonon
coupling. On the other hand, it is interesting to compare
MgB$_2$ with related isostructural compounds having much
lower transition temperatures or even showing no superconductivity at all. 
We choose here TaB$_2$ for which band structure 
calculations have been reported recently\\
\\
\\
\\
\\
\\
\\
\\
\\
\\
\\
\\
\\
\\
\\
\\
\vspace{.0cm}
\noindent
\cite{Rosner,Stein,Singh} and for which 
single crystals can be
produced \cite{Otani}.
 It should be noted that stochiometric TaB$_2$ is non-superconducting 
(at least down to 1.5 K \cite{Rosner,Leyarovska,Gasparov}) and another phase, 
unidentified yet, should be responsible for the observed superconductivity with
a $T_{c}$ of 9.5 K \cite{Kaczorowski}.

	\section{METHODS}

Very detailed insight into the electronic structure can be
gained from de Haas-van Alphen (dHvA) measurements. 
To the best of our knowledge there is as yet only one very recent 
measurement devoted to MgB$_2$ \cite{yelland} and
we are not aware of any dHvA-experiments for  TaB$_2$. With the aim to 
initiate possible experiments, we provided  calculated
dHvA-frequencies 
and effective masses for the two materials under
consideration \cite{saad}. From that data one can extract information about 
the Fermi 
velocities averaged over the extremal orbits. We discuss also the
possible enhancement of the effective masses by the electron-phonon
coupling which is expected to be quite different for the two
materials. Information about the Fermi velocities can furthermore be obtained
by means of the plasma frequencies detectable from the Drude part of the
optical conductivity. We compare also the plasma frequencies computed
for each Fermi surface sheet and find
a large amount of anisotropy in MgB$_2$ in difference to TaB$_2$.

The band structure calculations have been performed by means of two
computational codes: (i)
the fully relativistic augmented spherical waves method (ASW) 
\cite{Williams} and (ii) the scalar relativistic full potential local
orbital scheme (FPLO). In case (i) 
we used the von Barth-Hedin parametrization of the
exchange-correlation potential. A {\it k}-mesh of 148 special {\it k}-points in
the irreducible Brillouin zone part was chosen. 
For details of case (ii) see e.g.\ Ref.\ \cite{Rosnern,Rosner}.

Both compounds crystallize in
the hexagonal space group 
$P6/mmm$ with lattice constants $a=3.083$ \AA, $c=3.5208$ \AA\ and
$a=3.082$ \AA, $c=3.243$ \AA \ \cite{Kaczorowski} for MgB$_2$ and TaB$_2$,
respectively. The resulting ASW band structure for MgB$_2$ is similar
with those published in the literature
\cite{An}$^-$\cite{Rosnern} and it is shown for the sake of completeness 
in Fig.\ 1. 
The relativistic band
structure of TaB$_2$ is presented in Fig.\ 2. 
We have checked our results by comparing them with
the scalar relativistic full-potential local orbital (FPLO) method
repeating previous calculations (MgB$_2$ \cite{Fuchs}$^-$\cite{Rosnern},
TaB$_2$ \cite{Rosner}) and we found some differences for 
both compounds. In MgB$_2$ these differences are not 
\begin{figure}[Attachment1]
\vspace{-0.5cm}
\hspace{-0.15cm}
 \rotatebox{-90}{
\includegraphics[height=.34\textheight]
{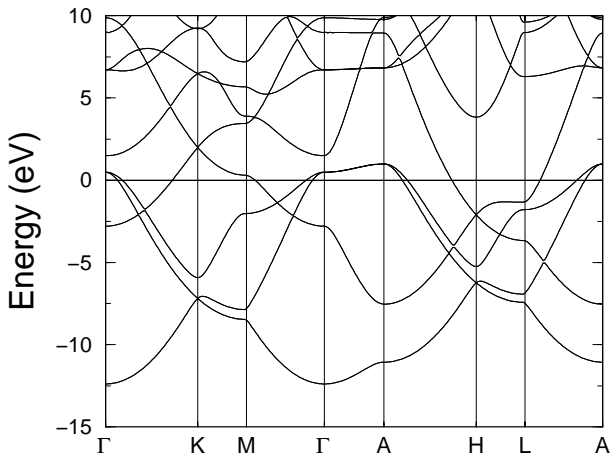}
}
\vspace{0.2cm}
\caption{The calculated energy bands of MgB$_2$. The Fermi energy
is at zero energy.}
\end{figure}
\noindent
\begin{figure}[Attachment2]
\vspace{-2cm}
 \hspace{-0.4cm}
\rotatebox{-90}{\centerline{
\includegraphics[height=.42\textheight]
{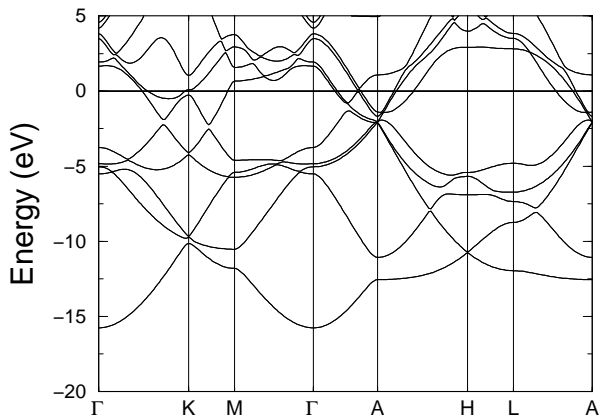}}}
\vspace{0.1cm}
\caption{The same as in Fig.\ 1 for TaB$_2$.}
\end{figure}

\noindent
 due to the spin-orbit coupling
in a
 fully relativistic
calculation. Instead 
  we are forced to ascribe these effects mainly to a slightly different 
crystal field potential in spherical or full-potential descriptions,
 which affects the relative position of boron
derived $\pi$ and $\sigma$ bands.
The spin-orbit coupling of the Ta 5$d$ states in
TaB$_2$ results in differences compared to scalar relativistic calculations, 
especially between $\Gamma$ and $A$ (compare Fig.\ 2
and Fig.\ 3 of Ref.\ \onlinecite{Rosner}). Although these differences should not
affect the absence of  
superconductivity in this compound, they lead to consequences for
the Fermi 
surface to be discussed below.
To predict the dHvA-frequencies $F$ for a magnetic field applied, say
along  the crystallographic $\vec{c}$ direction, we calculate the area $A$ of
any  
extremal orbit $\Omega$ according to the Onsager relation 
(written in cgs units)
\begin{equation}
F=\frac{\hbar c}{2{\pi}e}A=\frac{\hbar c}{2{\pi}e}{\int_{\Omega}}d^{2}k \quad .
\end{equation}
It is convenient to measure $F$
in units of the applied magnetic field and $A$ in units of squared
wave vectors, e.g.,\ in kT (1kT=10$^7$G) and the cross-section $A$ 
in 
nm$^{-2}$, respectively. Then Eq. (1) reads simply $$\hspace{2.6cm} F=0.104867 \times A 
\quad . \hspace{2.1cm} (1')$$ The effective 
mass of the corresponding orbit is given by
\begin{equation}
m=\frac{\hbar}{2{\pi}}{\oint}{\frac{dk}{|\vec v_{k}|}} , 
\hspace{1.3cm}            
\mbox{with} \hspace{1.3cm}   
\vec v_{k}=\frac{\partial{\epsilon_{k}}}{\partial{\vec{k}}}. 
\end{equation}
From the dHvA-frequency $F$ and the effective mass {\it m} one can introduce 
an effective Fermi velocity
\begin{equation}
v_{f}=\frac{\hbar}{m}{\sqrt{\frac{A}{\pi}}}
      =\frac{1}{m}{\sqrt{\frac{2{\hbar}eF}{c}}}, 
\end{equation} 
which corresponds to the hypothetical value for a circular orbit with
constant $|{\vec v_{\vec{k}}}|$. This expression is frequently used to 
estimate the Fermi velocity from dHvA data. Alternatively, one can also average
1/$|{\vec v_{k}}|$
around the circumference and define
\begin{equation}
 \tilde v_{f}=\frac{{\hbar}L}{2{\pi}m}=\frac{\hbar}{2{\pi}m}{\oint}{dk}.
\end{equation} 
Obviously, both velocities defined by Eqs.\ (3) and (4) coincide for a
circular orbit. For an orbit with an ideal hexagon shape $v_f/\tilde
v_f=
\sqrt{\pi/2\sqrt{3}}\approx 0.9523$ is estimated from Eqs.\ (2,3). 
Both of these quantities contain averages of point properties of the Fermi 
surface around a cyclotron orbit.

Information about the Fermi velocity is also contained in the
plasma frequencies $\omega_{p,\alpha}\,(\alpha=x(y)\ \mathrm{or}\ {\it z})$
\begin{equation}
(\omega_{p,\alpha})^{2}=\frac{e^{2}}{\epsilon_{\circ}}
\int{\frac{d^{3}k}{(2{\pi})^{3}}{
\delta{(\epsilon_{k}-E_{f})}}{(v_{k,\alpha})}^{2}},
\end{equation} 
where we will distinguish the contributions from different Fermi surface
sheets in the following (denoted by $\omega_{p,\alpha}^
{n}$). Analogously to Eq.\ (5), 
components $\alpha$ of the Fermi velocities $v_{\alpha}=\sqrt{<v^2_{\alpha}>_{FS}}$ 
can be introduced
which are 
 averaged over selected sheets or over the
whole Fermi surface (FS) 
\begin{equation}
(v_{\alpha})^{2}=
\int{\frac{d^{3}k}{(2{\pi})^{3}}{
\delta{(\epsilon_{k}-E_{f})}}{(v_{k,\alpha})}^{2}}.
\end{equation}
The ratio of these Fermi velocity components related to the most anisotropic 
$\sigma$-tubes with the strongest el-ph interaction at the same time  
can be used to describe
upper bounds for 
the anisotropy parameter $\gamma$ of the upper critical field $H_{c2}$,
$$\gamma=H_{c2}^{ab}/H_{c2}^{c} < \sqrt{<v_{f,xy}^2>_{tubes}/2<v_{f,z}^2>_{tubes}}
 \quad ,$$
 if that anisotropy is
the predominant one as suggested by several
 microscopic calculations \cite{Fuchs}.

\section{ Magnesium diboride}  
Let us first discuss the results for MgB$_2$. The band structure near 
the Fermi energy $E_f$ is depicted  in Fig.\ 1. The corresponding
dHvA-data for a magnetic field directed along the $c$-axis are shown 
in Table I. (Note that in our previous paper \cite{saad} the numbers for the
orbits 2 and 4 were mistakenly interchanged.)
The Fermi surface of    
\begin{figure}[Attachment3]
\hspace{-0.3cm}
 \includegraphics[height=0.14\textheight]{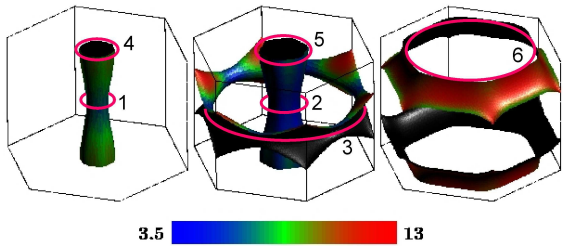}
 \vspace{0.2cm}
\caption{(Color) The calculated Fermi surface, distribution of Fermi
velocities,
 and extremal
orbits (pink lines, the 
numbers denote the orbits that are explained in the text)
(magnetic field $\parallel c$-axis) of MgB$_2$. The Fermi velocities
expressed by different colors (blue-slow, green-medium, read-fast) are
given in units of 10$^5$m/s as shown in the legend.}
\end{figure}
\vspace{0.3cm}
\begin{figure}[Attachment4]
\includegraphics[height=.12\textheight]
{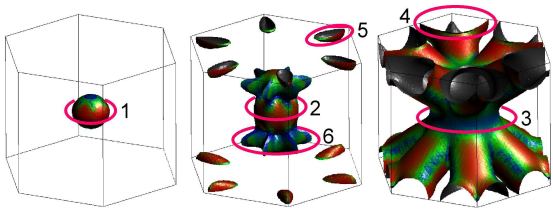}
\vspace{0.2cm}
\caption{(Color) The same as in Fig.\ 3 for TaB$_2$. Note that for
visibility purposes A is at the center of the hexagonal prism and
the $\Gamma$ point is the midpoint of the lower and the upper hexagon.} 
\end{figure}
\noindent
MgB$_2$
(see Fig.\ 3 )
consists of four sheets 
and there are 6 extremal orbits for a magnetic field in $\vec{c}$
direction. We will denote them by numbers 1 to 3 centered around $\Gamma$
and 4 to 6 centered around {\it A} with increasing area in each case. So one
can see that orbits 1 and 4 (2 and 5) belong to the smaller (larger)
tube, where the remaining orbits 3 and 6 are rather large and belong
to the ring-like parts of the Fermi surface. The
electrons on orbit 6 have the largest Fermi velocity and $v_{f}$ of 
orbit 3 is smaller compared to that of orbit 6 due to the larger effective
mass. The 
electrons on 
the tubes are slow in general, with the minimal velocity on orbit 2. The
ratio of Fermi velocities for the two different groups of electrons is 
roughly 2. In addition there is a rather different strength of the 
electron-phonon 
interaction on both types of Fermi surface sheets: strong on the tubes 
and weak on $\pi$-derived rings. This confirms 
the introduction of a two-band model proposed for MgB$_2$ first 
by Shulga {\it et al.}\cite{Shulga} and elaborated later on in 
Ref.\ \onlinecite{choi}.

The relative 
difference  $(v_{f}- \tilde v_{f})/v_{f}$ does not exceed 7 percent and it
is noteworthy only for those orbits which have a pronounced hexagonal
form than a
nearly  
circular one (for example orbit 3 is such a nearly circular orbit). According 
to the two-band
model \cite{Shulga} we would expect a different electron-phonon mass
renormalization (1+$\lambda$) for electrons on the tubes (with
$\lambda{\simeq 1.2\cdots1.5}$) and on the large orbits 3 and 6
($\lambda{\simeq 0.3}$). 
The comparison with the experimentally determined cross sections 
shows significant deviations from our L(S)DA predictions.
The microscopic reason for that unexpected deviation 
remains unclear at the moment. Standard electron-phonon interaction 
does not affect the cross-section. It might explain only the
renormalization of the Fermi velocities and of the masses.
The cross-sections
 might be affected by strong electron-electron interaction and/or 
the presence of electronic instabilities competing with
superconductivity
which might result in local changes (destruction) of the Fermi surface
by a partial dielectrization (gapping) from a charge density or spin
density wave \cite{hase}. The experiments yield a much narrower 
smaller tube than the L(S)DA calculation predict. While this already
indicates a discrepancy between theory and experiment, more serious
might be the fact that the wider, second tube is not observed in the
experiment at all. Would this fact be confirmed also for cleaner 
samples, i.e.\ the missing cross sections with  lower Fermi 
velocities could not be attributed to the residual impurity 
scattering\cite{yelland}, 
it would imply the failure of the L(S)DA to describe correctly 
MgB$_2$. 
 
We consider the elucidation of the responsible many-body effects
which are
missing in the standard LDA approach to MgB$_2$, i.e.\ treating it as 
a simple $sp$-metal, as a possible key problem
for our physical understanding of the electronic structure of MgB$_2$
and its remarkable superconductivity as well. Anyhow, 
in view of the great importance of the experimental dHvA data, the 
confirmation 
of a more or less unusual electronic structure in MgB$_2$ 
by other groups using different
single crystals would be highly desirable.
 
The experimental determination of the
effective masses by dHvA measurements and its comparison with the
calculated electronic values gives important information on
details of the electron-phonon coupling. If the assignment proposed 
in Ref.\ 19 is correct, strongly renormalized Fermi velocities by a 
factor of 2 to 3 should be explained. Comparing the calculated and measured
masses for the smaller tube, one arrives at a slightly larger local 
$\lambda \approx$ 1.27 compared 
with $\lambda \approx$ 1 obtained by Choi {\it et al.} \cite{choi}
within full anisotropic (harmonic) Eliashberg theory.

If the corresponding  coupling constant for the larger tube would be 
also enhanced, possible contributions from 
the coupling of a soft mode should be considered.
Otherwise it would be unclear, why for a non-enhanced
Coulomb repulsion (measured by the 
Coulomb pseudopotential $\mu^*$) or the absence of any other enhanced 
pair-breaking, the critical temperature is not 
higher than 40 K. In this context the recent experimental
observation of a soft-mode near 17 meV in Raman spectroscopy \cite{lampakis}, point
contact spectroscopy \cite{bobrov} as well as in neutron scattering 
\cite{murano} is of
considerable
interest. Especially with respect to the assignment of that boson-mode, since 
standard harmonic phonon calculations
are unable to explain the presence of such a low-lying  mode \cite{Kong}.
Notably, also from recent standard tunneling experiments \cite{dyachenko}
(unfortunately with a cut off at 
26 meV, i.e.\ just slightly above the low-frequency anomalies) 
 significant deviations
 from the standard harmonic phonon calculations have been reported.

The anisotropy of Fermi
velocities of MgB$_2$ is even more visible in our calculated plasma frequencies
(see Table II),
which are similar to the results of 
Refs.\ \onlinecite{Kong,Kortus,ravindran,singh} (see Table III).
In our previous calculation \cite{saad} a scaling factor of $c/a$ was
missing for $\omega_{p x(y)}$. Here we have corrected the present results
for that factor.
 It is another kind of anisotropy: the {\it z}-component of $v_{f}$
is about 6 times smaller than the  
{\it x(y)} components for the two tubes (sheets 1 and 2). The
anisotropy 
ratio $ v_{p,x(y)}^{n}/{v_{p,z}^{n}}\ \
(v_{\alpha}^{n}\propto{|\omega_{p,\alpha}^{n}}|)$ is much smaller for the  
sheets 3 and 4 and its average Fermi velocity is larger. The total
plasma frequency, however, shows only a small anisotropy of 1.04 between the
{\it z} 
and {\it x(y)} components. In Table III we compare the anisotropy in the 
components of the 
total plasma frequency as obtained by various groups. The obtained
numbers are generally in good agreement with one another.

The anisotropy of Fermi velocities is much less
pronounced in TaB$_2$ (Tables III and IV). Our fully relativistic Fermi
surface coincides almost with that published in \cite{Rosner} and consists of 
three sheets. The extremal orbits are characterized as follows (s.\
Fig.\ 3):
numbers 1 to 3 are centered around the {\it A} point and correspond to the
three different sheets. Orbit 4 is centered around {\it K} and orbit 5
corresponds to the small Fermi surface part in the neighborhood of
{\it K}. There is one maximal orbit (number 6) with nontrivial
value $k_{z}$ ($0.20615 \times 2\pi/a$). Since Ta is a  5$d$ element, 
the influence 
of the spin-orbit coupling is more pronounced in TaB$_2$ than in
MgB$_2$. With respect to the scalar relativistic calculations  
\cite{Rosner} we find two differences that affect the Fermi surface: 
First, a splitting of bands along the high symmetry direction
$\Gamma-{\it A}$, which leads to a lifting of band degeneracy at the Fermi
level. Second, also at the $K$-point a small band shift occurs. The
first spin-orbit modification of the Fermi surface makes one Fermi
 sheet disjointed along $\Gamma-{\it A}$. The calculated areas, 
effective masses and Fermi velocities of TaB$_2$ are collected in Table IV and
one  
can see that any kind of anisotropies of Fermi velocities is much smaller than
for 
 MgB$_2$. According to the calculation of the electron-phonon coupling
constants \cite{Rosner} we would expect a rather small renormalization 
of effective masses in TaB$_2$ ($\lambda{\simeq 0.05\cdots0.2}$) again in
difference to MgB$_2$. The plasma frequencies are collected in Table V. 
Since there are no analogous FS sheets to the two tubes of MgB$_2$
we do not obtain a corresponding anisotropy in
${\omega_{p,x(y)}^{n}}$ with respect to ${\omega_{p,z}^{n}}$.
This can be understood due to the loss of the quasi two-dimensionality of the
B $\sigma$-states in TaB$_2$ due to the hybridization with the Ta 5$d$
states. 

The hole Fermi sheet around A shown in the middle panel of Fig.\ 4 resembles 
a feature calculated within an augmented-plane-wave code for ZrB$_2$ 
\cite{ihara} The experimentally observed dHvA-frequencies of the central 
$\varepsilon$-orbit \cite{tanaka} (corresponding to orbit No.\ 2 in Fig.\ 4) 
near 1.8 kT and 1.5 kT for the similar system TiB$_2$ \cite{ishizawa}, 
respectively,
are  significantly 
smaller than 3.48 kT in our FPLO-calculation for 
TaB$_2$. The opposite behaviour is observed for the $\mu-$orbit (orbit No.\ 6
in Fig.\ 4) where the experimental values for ZrB$_2$ and TiB$_2$ read 2.46 kT
and 3.44 kT compared with our FPLO calculated 3.44 kT for TaB$_2$.
 Further theoretical and experimental 
 studies should be awaited to answer the important question
whether there are significant deviations even from full-potential 
L(S)DA-calculations only for the peculiar MgB$_2$ 
or also for other diborides with weak electron-phonon interaction.

\section{Summary and Conclusions}

In summary, we have calculated the dHvA-frequencies and effective
masses as well as the plasma frequencies of   MgB$_2$ and TaB$_2$ which would
be  
worth to be investigated experimentally in detail.
First experimental de Haas-van Alphen 
data for MgB$_2$ point to significant deviations. 
 The comparison of the experimental
masses with L(S)DA calculation could give important information on the
electron-phonon coupling. According to Refs.\ \cite{An,Rosnern,Rosner} we
expect 
remarkable differences between 
 MgB$_2$ and  TaB$_2$: Large enhancement for one group of electrons in 
 MgB$_2$ and a small enhancement for the other electrons in MgB$_2$
 as well as in TaB$_2$. We found a remarkable anisotropy of Fermi
velocities in superconducting  MgB$_2$ which is in difference to the
more isotropic behavior of non-superconducting TaB$_2$.\\
{\it Note added.}After finishing the present 
second version of our 
work we have learnt on 
a closely related work by Mazin and Kortus\cite{mazin2} who also stress 
the necessity of full-potential calculations for MgB$_2$ 
and arrive essentially at very
close numerical values for the cross sections and the masses 
under consideration 
(see Table I). Remaining slight numerical 
differences, comparison with other full-potential bandstructure codes, 
 and somewhat different physical interpretation will be discussed elsewhere
 \cite{rosner3}. 

\begin{acknowledgments}

We thank S.V.\ Shulga, I.\ Mazin, and A.\ Carrington for critical 
remarks and discussions.
We are indebted to H.\ Eschrig, M.\ Richter, V.D.P.\ Servedio, 
W.\ Pickett and J.\ An for
discussions. This work 
was supported by the Egyptian Ministry of Higher Education and
Scientific Research and by   
the DAAD (individual grant to H.R.), ONR Grant No. N00017-1-0956 and the DFG,
SFB 463.
\vspace{0.5cm}
\begin{center}
\begin{tabular}{l}
\hline{}
\phantom{aaaaaaaaaaaaaaaaaaaaaaaaaaaaaaaa}\\
\end{tabular}
\end{center}

$^{1}$ On leave from Faculty of Science, Menoufia University, Shebin El-Kom,
Egypt 

\end{acknowledgments}


\newpage
\onecolumn
\begin{table}
\caption{The LSDA de Haas-van Alphen data of MgB$_2$ 
(magnetic field $\parallel c-$ 
axis). Available  
experimental values according to the assignment of Ref.\ 19 have been
added for the sake of comparison. Calculated quantitities without
explicit label have been computed within the ASW scheme.}
\begin{tabular}{crrrrccccrrr} 
orbit&${F_{ASW}}$&$F_{FPLO}$&${F_{exp}}$&$A/A_{BZ}$&$m/m_{\circ}$&$m_{exp}/m_{\circ}$&
${v_{f}}$&$v_{f,exp}$&${\tilde v_{f}}$\\ [4pt]
 \hline
&[kT]&[kT]&[KT]&\%&& &[10$^5$m/s]&[10$^5$m/s]&[10$^5$m/s]\\ [4pt]
\hline
1  &  0.99  &0.79&0.54&   1.98 & 0.25 &0.57&  7.9 $$ &3.2$$& 8.5 $$\\ [4pt]
2  &  2.16  &1.63&    &  4.3$$ & 0.55 & &  5.4 $$ & $$& 5.8 $$\\ [4pt] 
3  & 32.88  &34.2&     &65.6 & 1.70 &    &  6.8 $$ && 7.2 $$\\ [4pt]
4  &  2.31  &1.84&1.53     & 4.6 & 0.31 &  0.70  &  9.8 $$ &3.3& 10.5 $$\\[4pt]
5  &  4.31 &3.44&     & 8.6$$ & 0.61 &    &  6.8 $$ && 7.3 $$\\ [4pt]
6  & 30.20 &30.6    &     &60.2$$ & 0.92 &    &  12. $$ && 12.8 $$\\ [4pt] 
\end{tabular}
\end{table}
\twocolumn
\begin{table}
\caption{
Squared plasma frequencies of MgB$_2$ in 
[eV$^{2}$]. One eV
for the plasma frequencies corresponds to 80.4$\cdot $ km/s.} 
\begin{tabular}{ccrc} 
FS sheet&${\omega_{p, x(y)}^{2}} $&$\omega_{p,z}^{2} $&
$v_{x(y)}/v_{z}$ \\ [4pt]
\hline
1 (tube) &9.639&0.242 &6.31\\ [4pt]
2 (tube) &9.444&0.323 &5.40\\ [4pt]
3 (ring) &8.620&22.053&0.63\\ [4pt]
4 (ring) &23.042&24.6698&0.97\\[4pt]
\hline
All      &50.75&47.290&1.04\\ [4pt]
\end{tabular}
 \end{table}
   
  \begin{table}
  \caption{Comparison of several calculated results for the two
components of the total plasma frequency $\omega_{p x}$ and
$\omega_{pz}$ of MgB$_2$ (in eV).}
\begin{tabular}{rrc}
${\omega_{p, x(y)}} $&$\omega_{p,z} $&
Reference \\ 
\hline
7.12&6.88&this work\\
7.21&6.87&Ref.\ 3\\
7.02&6.68&Ref.\ 4\\
7.13 &6.72 &Ref.\ 28\\
7.04&6.77&Ref.\ 29\\ 
\end{tabular}
\label{tab3}
\end{table}

\begin{table}
\caption{
Calculated de Haas-van Alphen data of TaB$_2$.}
\begin{tabular}{crrrcccrr}    
orbit&${F_{ASW}}$&${F_{FPLO}}$&${A_{ASW}/A_{BZ}}$&${m}/{m_{\circ}}$&
$v_{f}$&$\tilde{v}_{f}$\\
\hline
&[KT] &[KT]&&&[km/s]&[km/s]\\ 
\\ \hline
1  &  2.08  & 2.50  & 0.041 & 0.221 &  1320  & 1410  \\ 
2  &  2.99  & 3.48  & 0.059 & 0.356 &  979  & 1050 \\ 
3  &  8.40  & 8.31  & 0.167&0.677 &  864  & 926 \\
4  & 13.26  & 13.84 &0.264 & 0.558 &  1320  & 1410 \\ 
5  &  1.05  & 1.00  & 0.021 & 0.236 &  869  & 891 \\
6  &  1.15  & 3.44  & 0.023 & 0.632 &  339  & 365 \\ 
\end{tabular}
\end{table}
  \begin{table}
  \caption{Squared 
plasma frequencies $\omega_{p x}$,
$\omega_{pz}$, and the anisotropy of averaged 
Fermi velocities for TaB$_2$ (in eV$^2$).}
\begin{tabular}{crrc}
FS sheet&${\omega_{p, x(y)}} $&$\omega_{p,z} $&
$v{x(y)}/v_{z}$ \\ 
\hline
1&68.125&84.779&0.90\\
2&15.149&17.049&0.94\\
3&3.243 &3.254 &1.00\\
\hline
All&86.52&105.08&0.91\\ 
\end{tabular}
\label{tab5}
\end{table}

\end{document}